\documentclass[prb,preprint]{revtex4}
\usepackage{amsmath}
\usepackage{graphicx}
\begin{document}
\title{Search for chaotic behavior in a flapping flag} 
\author{J. O. McCaslin and P. R. Broussard} 
\affiliation{Covenant College, Lookout Mountain, Georgia 30750}

\begin{abstract}
We measured the correlation of the times between successive flaps of a flag for a variety of wind speeds and found no evidence of low dimensional chaotic behavior in the return maps of these times. We instead observed what is best modeled as random times determined by an exponential distribution. This study was done as an undergraduate experiment and illustrates the differences between low dimensional chaotic and possibly higher dimensional chaotic systems.
\end{abstract}

\maketitle

\section{Introduction}

Low dimensional chaotic behavior has been seen in many dynamical systems. The study of a driven pendulum,\cite{Berdahl} dripping faucets,\cite{Dreyer} and many other systems have illustrated the transition from periodic to chaotic behavior. The dynamics of a flapping flag have also been claimed to be chaotic,\cite{Lorenz,Vernier} but our attempt to find experimental data on flapping flags was not successful. There has been extensive work on the onset of flapping and flutter for paper and fabrics in air,\cite{Watanabe} as well as filaments in a two-dimensional soap film\cite{Zhang1} and heavy flags in water.\cite{Zhang2} However, none of these studies has characterized the dynamics of the systems studied as chaotic. The goal of this paper is to determine if low dimensional chaotic dynamics is revealed by looking at return maps of the time between successive flaps of a flag.

\section{Experimental Setup and Data Analysis}

\begin{figure}[h!]
\begin{center}
\includegraphics*[width=3.5in]{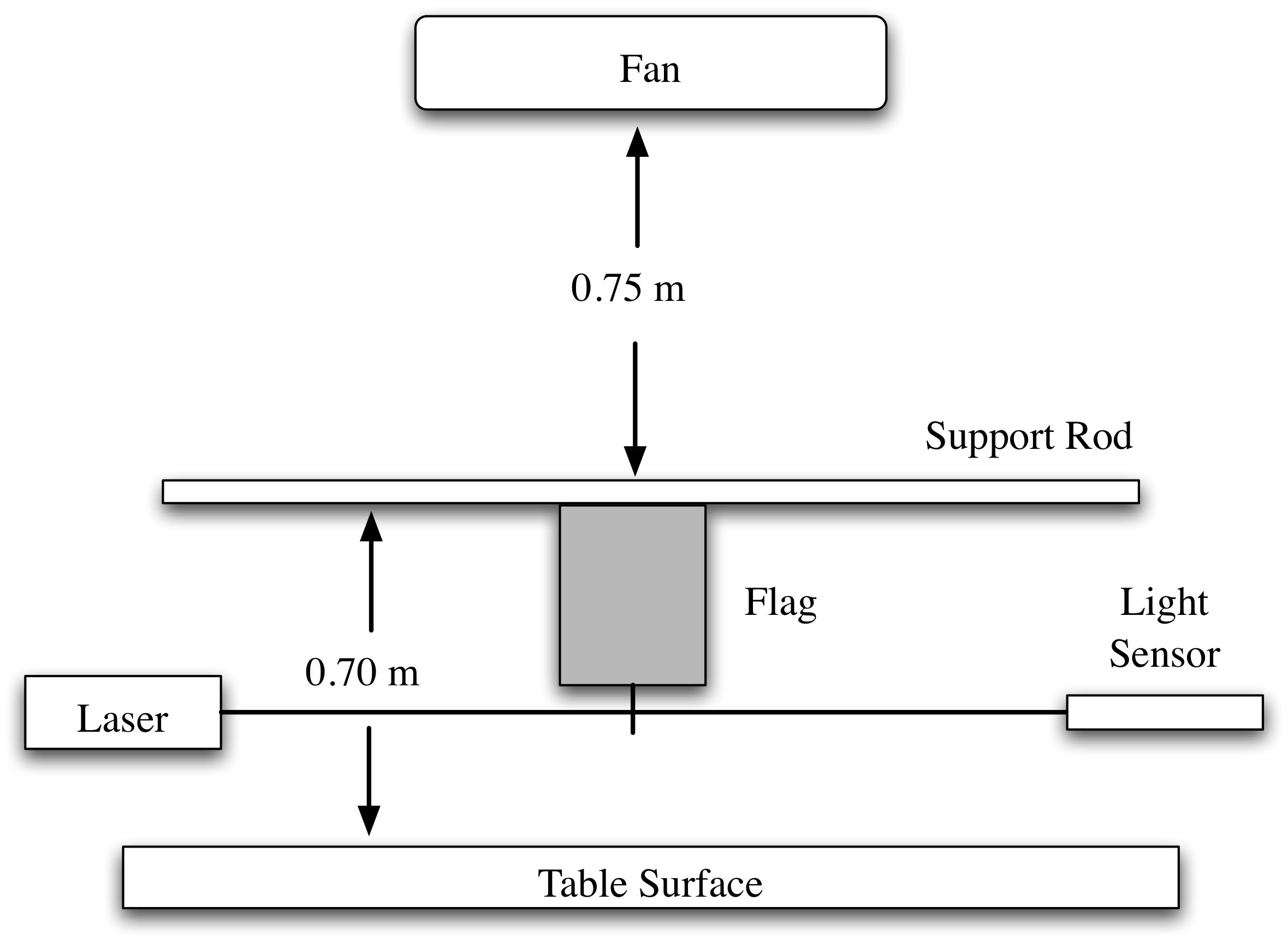}
\end{center}
\caption{Schematic of the experimental setup.}
\label{Setup}
\end{figure}

A schematic of the experiment is shown in Fig.~\ref{Setup}. In order to not deal with changes in the flag's aspect ratio due to gravity, we let the flag hang vertically and have the wind directed vertically. The wind was produced from a square home box fan (approximately 0.5\,m on a side) located approximately 0.75\,m above the flag support rod, in order to have a more uniform distribution of wind. The flag was approximately 0.35\,m long and 0.25\,m wide, and the support for the flag was approximately 0.70\,m above the supporting table so as to create a space for the air to move. Wind speed was controlled by using a variac to give a variable voltage to the fan. The voltage was varied from 30 to 120\,V in steps of 10\,V. The actual wind speeds were measured using a Skywatch handheld windspeed detector. Ten measurements of wind speed were taken for each voltage and then averaged. The relation between the wind speed and the voltage was approximately linear and is given in Table~\ref{Speeds}. 

\begin{figure}[h!]
\begin{center}
\includegraphics*[width=2.75in]{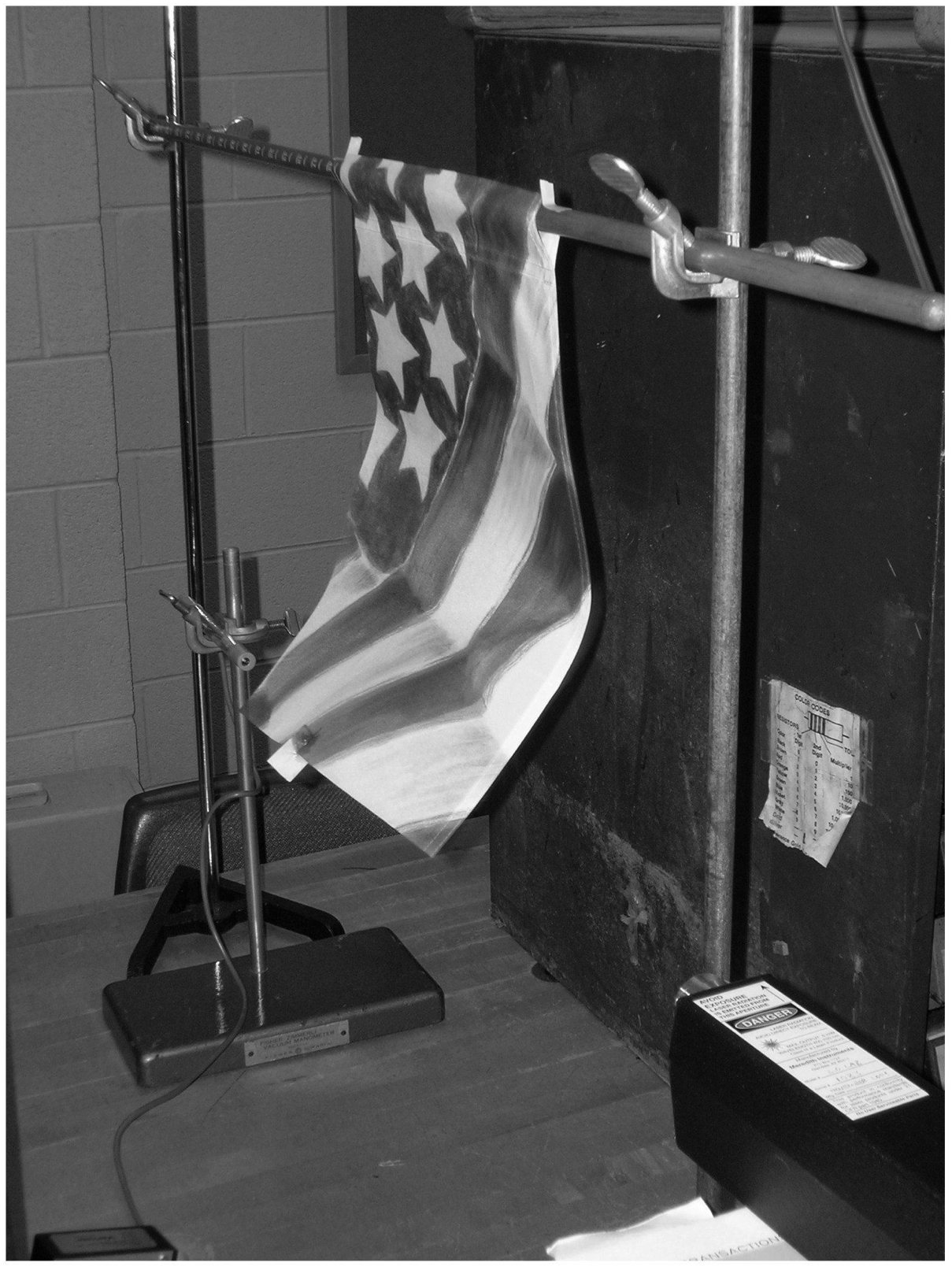}
\end{center}
\caption{Flag breaking beam.}
\label{Flag}
\end{figure}

\begin{figure}[h!]
\begin{center}
\includegraphics*[width=3.5in]{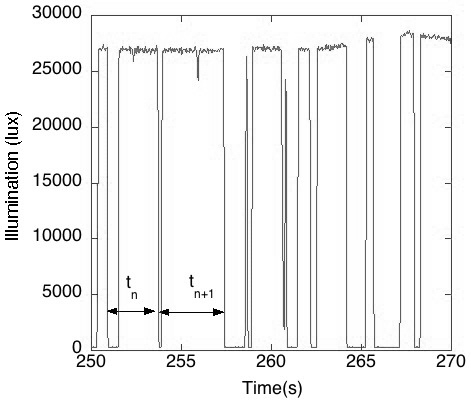}
\end{center}
\caption{Example illumination change as the flag flaps. This plot is for a voltage setting of 30\,V, or a wind speed of approximately 0.40\,m/s. The plot shows two example time intervals between successive flaps.}
\label{collection}
\end{figure}

The flapping of the flag was measured using a tab glued to the edge of the flag, which would break the beam from a HeNe laser (shown in Fig.~\ref{Flag}) as suggested in Ref.~\onlinecite{Vernier}. The laser's illumination was measured with a Vernier light sensor and Logger Pro software running on an inexpensive laptop. The rate of collecting illumination data was determined by making data collection runs at lower collection rates until the sensor began to ``miss'' the flapping. The data collection rate was set to 35 samples/s, and the data was collected for 15\,min for each wind speed. An example of the illumination versus time is given in Fig.~\ref{collection}.

To analyze the data, a program\cite{program} was used to examine the illumination versus time series to look for the time intervals between the flag flapping, defined in our case as the time between drops in intensity. The code outputs a list of time intervals, denoted as $t_{1}, t_{2}, \ldots, t_{N}$ for each wind speed. These data were then used to create plots of the time intervals versus their order as well as return maps, where the correlation between time intervals are studied.

\section{Experimental results}

\begin{figure}[h!]
\begin{center}
\includegraphics*[width=3.5in]{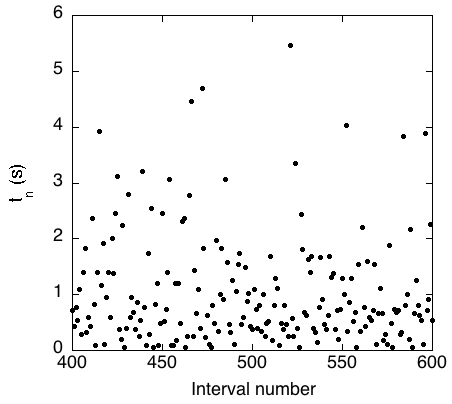}
\end{center}
\caption{The time between successive flaps versus the sequence iteration. This data was collected at 120\,V.}
\label{interval}
\end{figure}

\begin{figure}[h!]
\begin{center}
\includegraphics*[width=3.5in]{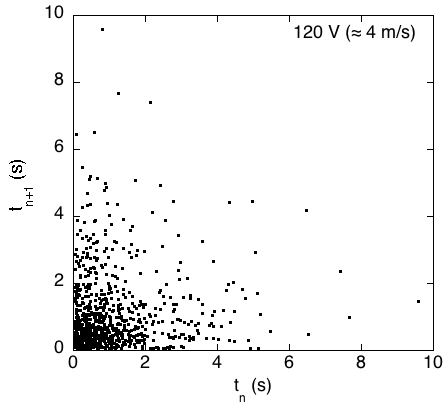}
\end{center}
\caption{Return map for the data collected at 120\,V.}
\label{returnmap120}
\end{figure}

\begin{figure}[h!]
\begin{center}
\includegraphics*[width=7in]{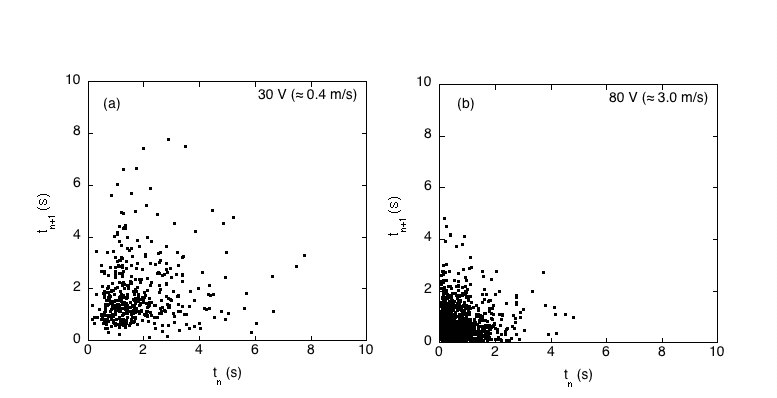}
\end{center}
\caption{Return map for the data collected at (a) 30\,V and (b) 80\,V. }
\label{comp}
\end{figure}

An example partial time series is shown in Fig.~\ref{interval}. The times are not uniformly distributed, but are more clustered around $t=0$. If we plot the time intervals in a standard two-dimensional return map ($t_{n+1}$ versus $t_{n}$) as in Fig.~\ref{returnmap120}, we see that there is no apparent pattern in the plot, as would be expected if the dynamics controlling the flapping exhibited low dimensional chaos, such as seen in dripping faucet times\cite{Dreyer} or the times between swings of a chaotic pendulum.\cite{Berdahl}  Data was collected for voltages from 30\,V to 120\,V (in steps of 10\,V) (wind speeds from just under 0.5\,m/s to over 4\,m/s).  All the return maps showed similar behavior, as seen in Fig.~\ref{comp} for the 30\,V and 80\,V data sets. The 30\,V data exhibits a clustering of points not around $t=0$, but instead near $(t_{n},t_{n+1})=(1,1)$. This behavior is partly due to the reduction of short time scales in the flapping of the flag. As the voltage decreases and the wind speed lowers, there are fewer and fewer rapid flapping events. In addition, there is a change in the distribution of times as we will discuss. The overall behavior of the return maps (higher density of points at low times, lower density for high times, no apparent structure) is unchanged.

\begin{figure}[h!]
\begin{center}
\includegraphics*[width=3.5in]{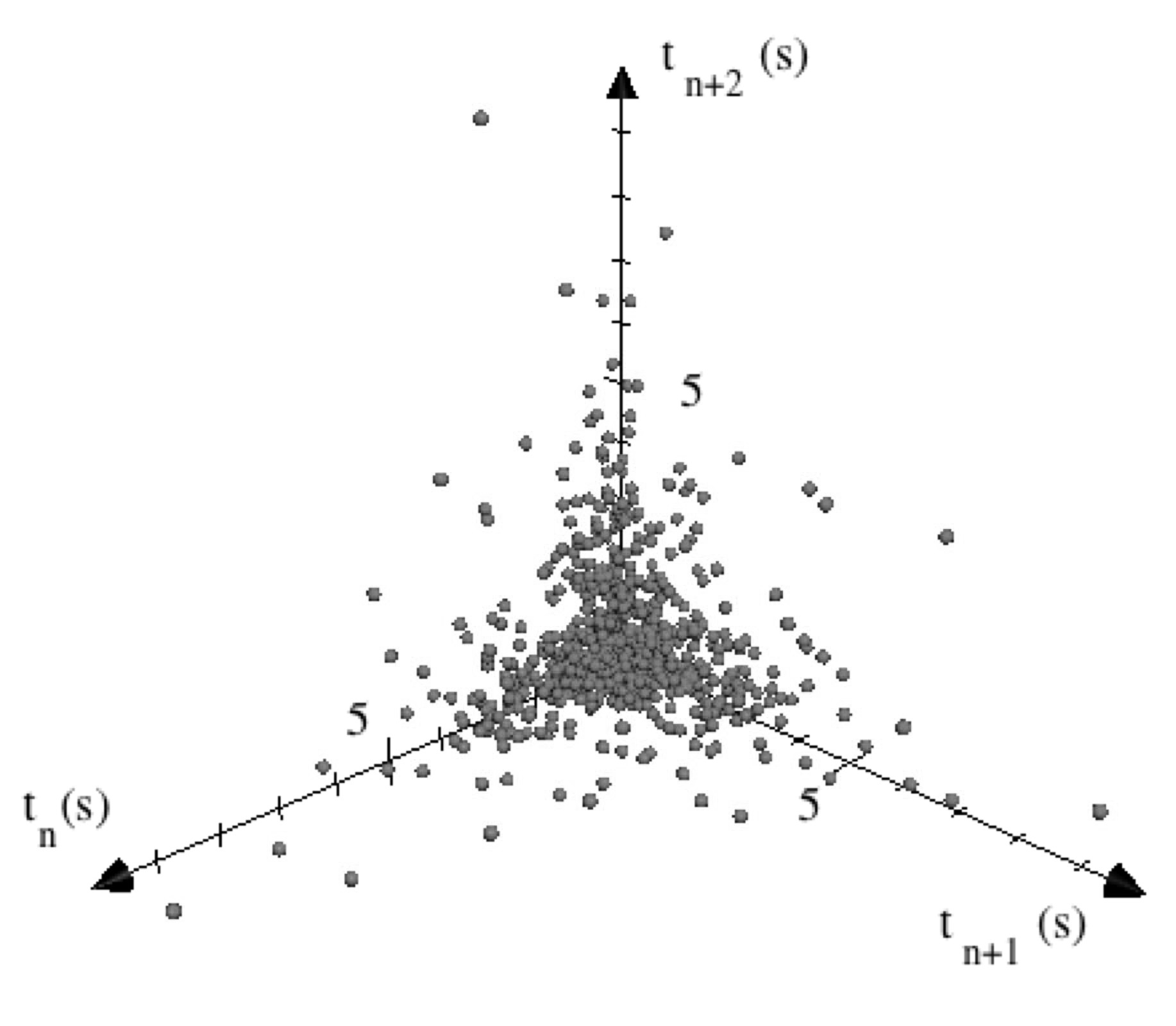}
\end{center}
\caption{Three-dimensional return map for the data collected at 120\,V. The tick marks on the axis correspond to 1\,s.}
\label{3D}
\end{figure}

As mentioned, the lack of structure in the return maps is evidence for the lack of low dimensional chaotic dynamics in this experiment. There is a possibility that the chaotic behavior could be higher dimensional and not revealed in a simple two-dimensional return map. To investigate this possibility, three-dimensional return maps were made with $(x,y,z)=(t_{n},t_{n+1},t_{n+2})$ as shown in Fig.~\ref{3D} for the 120\,V data set. Again, there is no apparent structure observed in this return map. We observe similar three-dimensional return maps for the other wind speeds. 

\begin{figure}[h!]
\begin{center}
\includegraphics*[width=3.5in]{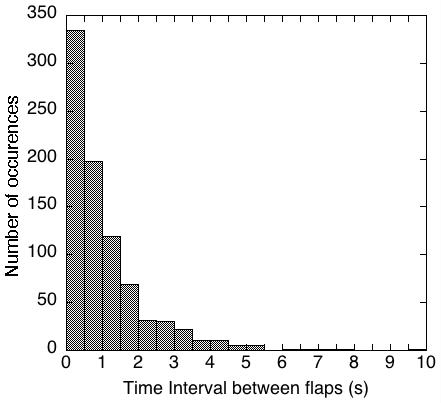}
\end{center}
\caption{Histogram of times between flaps for the 120\,V run.}
\label{hist}
\end{figure}

The return maps for voltages above 50\,V (or windspeed greater than 0.8\,m/s) show strong clustering near the origin, implying that shorter time intervals are more probable than longer ones. (Even for the lower values of windspeed where there seems to be a minimum time between flaps, there is still clustering around the lowest times.) For the three-dimensional return maps, the points are also densest near the origin, and then next along the axes, with fewer points along the diagonals. Although there are differences in the time scaling, the lack of any apparent structure in the return maps motivated us to look at other possible explanations. Figure~\ref{hist} shows that a histogram of the times follows an exponential distribution. The time constant is order 1\,s, with no systematic dependence on the wind speed. For the lowest windspeed measured, the histogram shows a peak at times greater than zero (1.2\,s for the 30\,V data) which could be due to the natural period of the flag (estimated to be 0.8\,s). For low wind speeds the dynamics could be due to the natural periodic motion of the flag. However, we did not observe any indication of periodic motion in the low wind speed return maps. As the windspeed is increased, the peak in the histograms decreases and for 50\,V and higher the peak is located at time equal zero.

\begin{figure}[h!]
\begin{center}
\includegraphics*[width=7in]{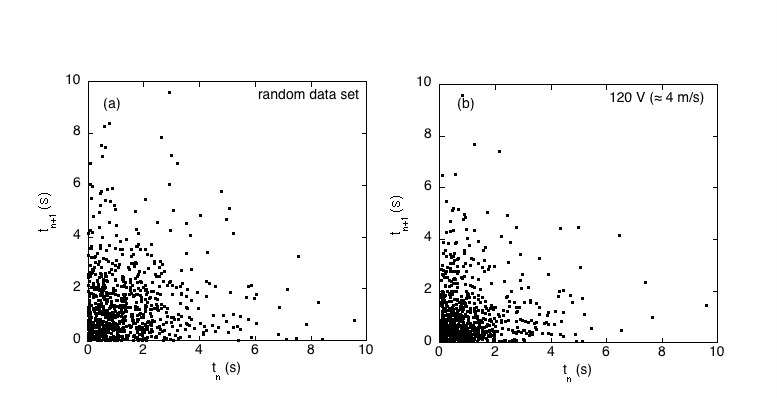}
\end{center}
\caption{Comparison of the return maps for times generated by an exponential probability (a) with the figure of data collected at 120\,V as shown in Fig.~\ref{returnmap120} (b).  The random data was scaled to match the 120\,V run in the number of points and range of times.}
\label{randomap}
\end{figure}

To compare the observed histogram order, we generated a random distribution of time intervals with an exponential probability. A set of $N$ uniform random numbers $x_i$ were generated in the interval (0,1), where $N$ was equal to the number of time intervals in one of our data sets. Exponentially distributed numbers $y_i$ were generated from the $x_i$ by the relation $y(i)=-\ln(x_i)$\cite{numeric} and then scaled to match the range of collected times for the data set considered. This set was then used to produce a simulated return map that would be expected if the time intervals between flaps are uncorrelated and occur with an exponential probability. Such a return map is shown in Fig.~\ref{randomap} along with an actual return map.

By comparing the randomly generated return map to the return map for 120\,V, it is clear that the two show very similar behavior, namely, strong concentration of points for small times and lower probability for large time intervals. We did not find evidence of low dimensional chaotic behavior for the time intervals between flaps, but instead data that resembles uncorrelated times with an exponential probability distribution, which would be equivalent to an infinite dimensional dynamic system. It is possible that we did not take data over a wide enough range of wind speeds. Because we did not observe any periodic flutter, such as seen in the heavy flag in the water experiment,\cite{Zhang2} the system should be in a non-linear dominated regime (even though we did see behavior at low windspeeds that might have been due to the natural period of the flag). We must also consider that our data may need to be analyzed at a higher dimensional level (looking at projections into 3D space of higher dimensional return maps) to verify that there is no chaotic nature to the results.  However, we can conclude that low dimensional chaos cannot be easily observed in a flapping flag experiment.\cite{Vernier}

\begin{acknowledgments}
The authors would like to gratefully acknowledge Dr.\ John Hunt for assistance with Java programing and Dr.\ Donald Petcher for his help in the random number analysis. In addition, the authors express their thanks to the anonymous referees who made many helpful suggestions for improving this paper.
\end{acknowledgments}

\newpage

\section*{Tables}

\begin{table}[h]
\caption{Voltage on fan and corresponding windspeed measured at the flag position.}
\begin{center}
\begin{tabular}{|c|c|}
\hline 
Voltage\,(V) & Windspeed\,(m/s) \\
\hline
$30 \pm 0.5$ & $0.40 \pm 0.23$ \\
\hline
$40 \pm 0.5$ & $0.52 \pm 0.16$ \\
\hline
$ 50 \pm 0.5$ & $0.83 \pm 0.24$ \\
\hline
$60 \pm 0.5$ & $1.33 \pm 0.28$ \\
\hline
$70 \pm 0.5$ & $2.36 \pm 0.64$ \\
\hline
$80 \pm 0.5$ & $2.98 \pm 0.64$ \\
\hline
$90 \pm 0.5$ & $3.16 \pm 0.77$ \\
\hline
$100 \pm 0.5$ & $3.52 \pm 0.81$ \\
\hline
$110 \pm 0.5$ & $3.87 \pm 0.74$ \\
\hline
$120 \pm 0.5$ & $4.07 \pm 0.76$ \\
\hline
\end{tabular}
\end{center}
\label{Speeds}
\end{table}

\newpage

\section*{Figure Captions}

\end{document}